\documentclass{v16nufact-oy}

\def\be{\begin{equation}}
\def\ee{\end{equation}}
\def\bea{\begin{eqnarray}}
\def\eea{\end{eqnarray}}

\newcommand{\Photo}{}

\begin{document}
\vspace*{4cm}
\title{Complementarity Between Hyperkamiokande and DUNE}

\author{OSAMU YASUDA \footnote{Speaker}, SHINYA FUKASAWA and MONOJIT GHOSH}
\address{Department of Physics, Tokyo Metropolitan University, Hachioji, Tokyo 192-0397,  Japan}

\maketitle\abstracts{
In this talk we present our results on the sensitivity to the neutrino mass hierarchy, the octant of the mixing angle 
and the CP phase in the future long baseline
experiments T2HK and DUNE as well as in the atmospheric neutrino
observation at Hyperkamiokande (HK).
}

\section{Introduction}
Thanks to the neutrino experiments in the last two decades,
the values of the three mixing angles and the values of the mass squared differences are now determined
in the three flavor mixing framework to some precision.
The unknown quantities at present are: (i) the mass hierarchy or the sign of $\Delta m_{31}^2$ (NH: normal hierarchy i.e., $\Delta m_{31}^2 > 0$ or IH: 
inverted hierarchy i.e., $\Delta m_{31}^2 > 0$),
(ii) the octant of $\theta_{23}$ (LO: lower octant i.e., $\theta_{23} < 45^\circ$ or HO: higher octant i.e., $\theta_{23} > 45^\circ$) and 
(iii) the CP phase $\delta_{CP}$.
These unknown quantities are expected to be determined in the future
experiments, such as T2HK\,\cite{Abe:2014oxa}, DUNE\,\cite{Acciarri:2015uup} and the atmospheric neutrino measurement at HK\,\cite{Abe:2011ts}, etc.
The T2HK experiment is an upgrade of the ongoing T2K experiment which will use a detector to have a volume
almost 25 times larger than the existing T2K detector. HK is the atmospheric counterpart of the T2HK experiment. On the other hand DUNE is a high statistics beam based
experiment to use high beam power, large detector volume and longer baseline.
In this talk
we study for the first time the joint sensitivity of the long-baseline experiments T2HK, DUNE and the atmospheric experiment HK in determining the 
remaining unknowns in neutrino oscillation sector.\footnote{
This talk is based on Ref.\,\refcite{Fukasawa:2016yue}.}
In this work we study: (i) the sensitivity of the T2HK, HK and DUNE experiments, (ii) the synergy between the T2HK and HK experiments to resolve the parameter degeneracy in the neutrino oscillation, 
(iii) how far the sensitivities in determining hierarchy, octant and CP can be stretched when all these three powerful experiments are combined together and 
(iv) the precision measurements of $\theta_{23}$, $\delta_{CP}$ and $\Delta m_{31}^2$ of this setup.

\section{Experimental Specification}

For our analysis of T2HK, we took the specification from Ref.\,\refcite{Abe:2014oxa}. We have considered a fiducial volume of 560 kt and a total exposure of $15.6 \times 10^{21}$ protons on target (pot) running in equal 
neutrino-antineutrino mode. The systematics are taken as an overall normalization error of 2\% (5\%) for appearance channel and
0.1\% (0.1\%) for disappearance channel corresponding to signal (background). 
For DUNE we have taken the details of the configuration from Ref.\,\refcite{Acciarri:2015uup}. 
A flux corresponding to 1.2 MW beam power is considered in our analysis. We have considered 5 year runtime in neutrino mode and 5 year
runtime in antineutrino mode. The detector volume is taken to be 35 kt. For systematics we have taken an overall normalization error of 2\% (10\%) for appearance channel and
5\% (15\%) for disappearance channel corresponding to signal (background). 
For HK we have taken the details from Ref.\,\refcite{Abe:2011ts}. A 560 kt water Cerenkov running for 10 year is considered. The systematics are same as used in Ref.\,\refcite{Fukasawa:2015jaa}.

\section{Combined Sensitivity of T2HK, HK and DUNE for determining the unknown parameters}

\subsection {Hierarchy}

\begin{figure}
\hspace{-0.05 in}
\includegraphics[width=0.35\linewidth]{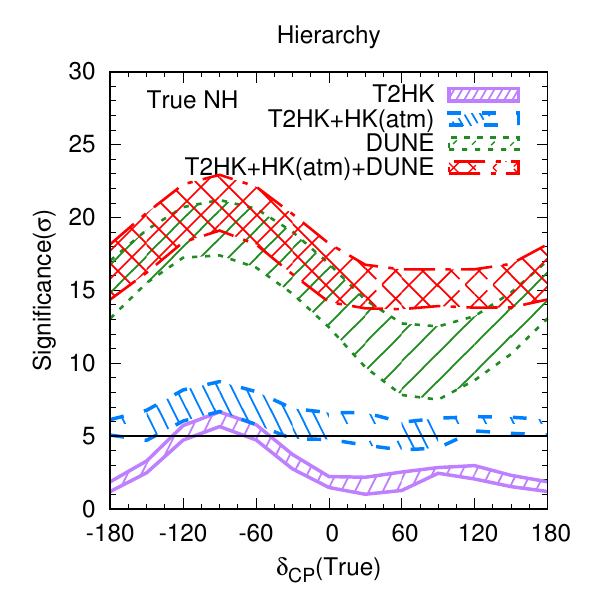}
\hspace{-0.25 in}
\includegraphics[width=0.35\linewidth]{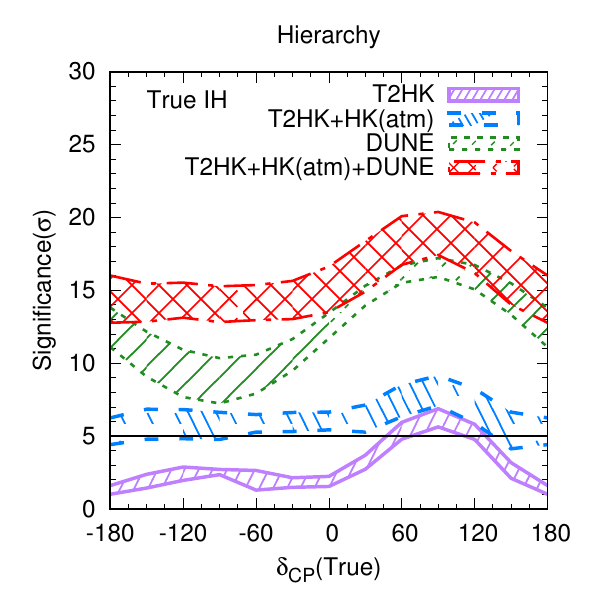}
\hspace{-0.25 in}
\includegraphics[width=0.35\linewidth]{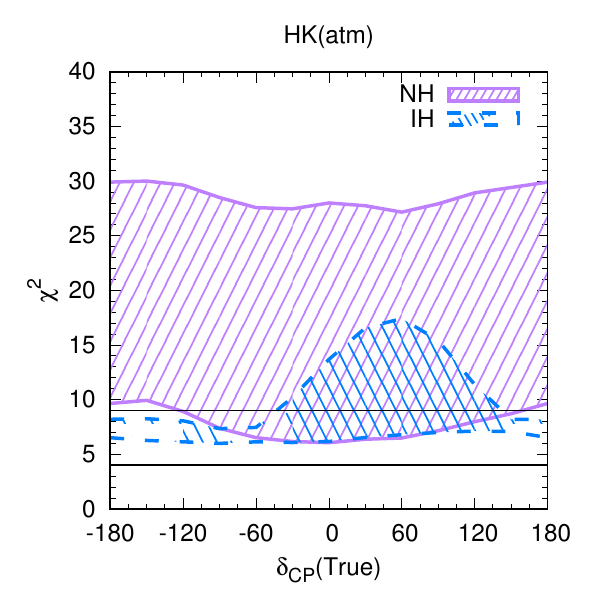} \\
%\hspace{-0.5 in}
\includegraphics[width=0.5\linewidth]{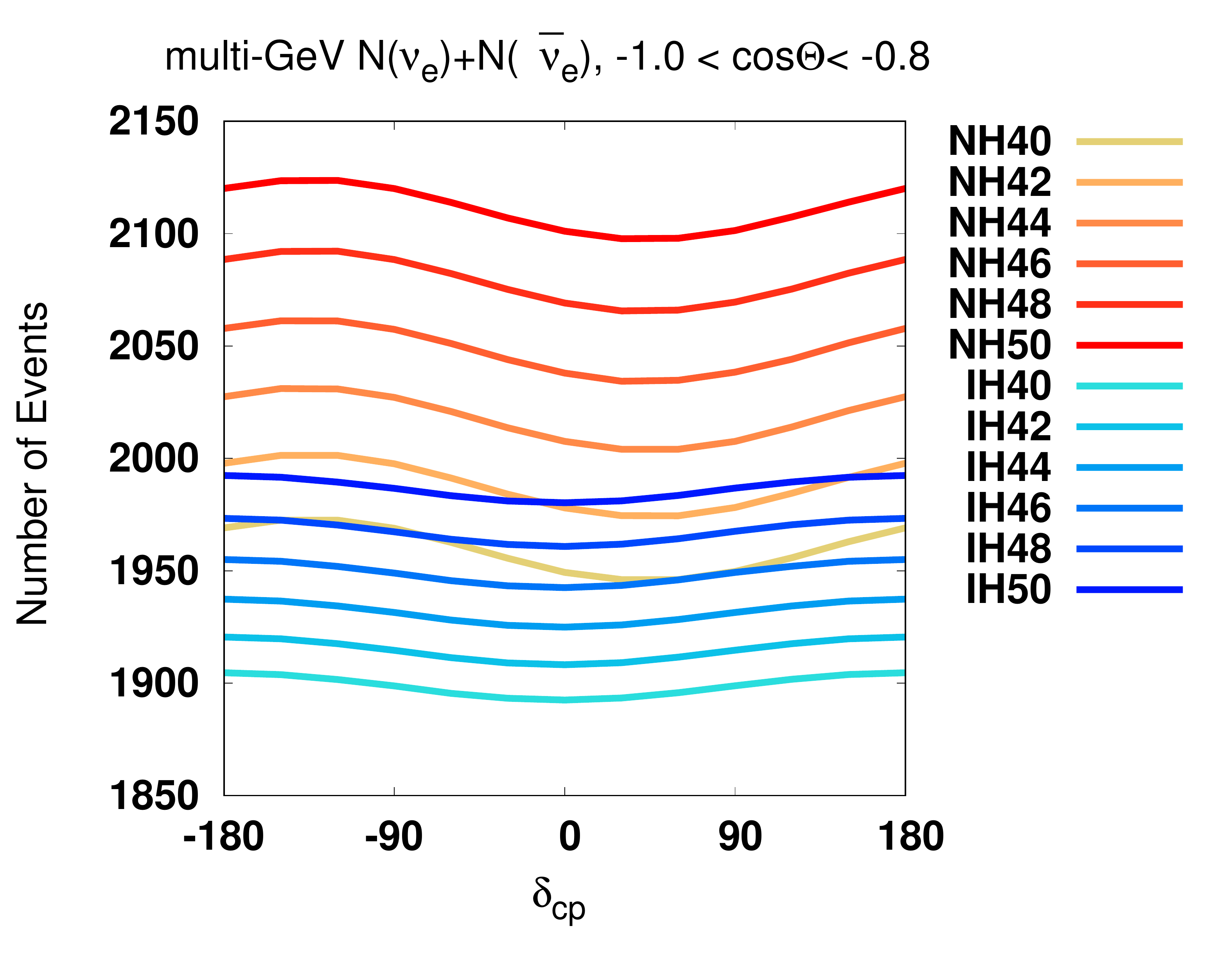}
%\hspace{-0.4 in}
\includegraphics[width=0.5\linewidth]{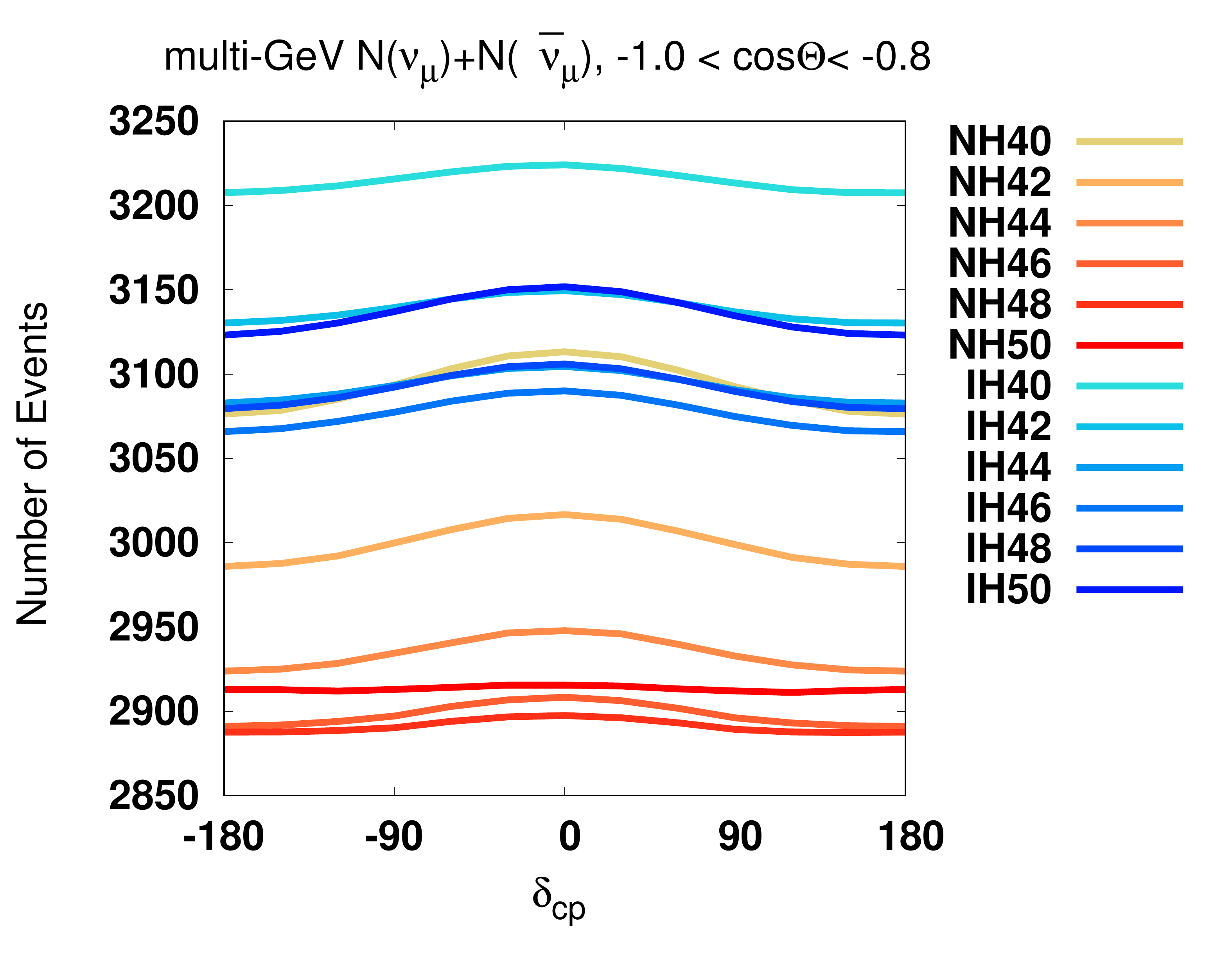}
\caption{Hierarchy sensitivity}
\label{fig1}
\end{figure}

We have calculated the hierarchy $\chi^2$ by taking the true hierarchy in the simulated data and wrong hierarchy in theory. We have marginalized over $\theta_{23}$ from $38^\circ$ - $52^\circ$ and
$\delta_{CP}$ from $-180^\circ$ to $+180^\circ$. We have given our results in Fig \ref{fig1}.
From the left and middle panels of the upper row we see that for T2HK the hierarchy sensitivity is poor in the unfavorable parameter space i.e., $0^\circ < \delta_{CP} < 180^\circ$ for NH and
 $-180^\circ < \delta_{CP} < 0^\circ$ for IH. This is due to the presence of hierarchy-$\delta_{CP}$ degeneracy.
 In these plots the width of the bands is due to the variation of $\theta_{23}$ from $40^\circ$ to $50^\circ$. If we add the HK data to T2HK then there is an improvement in the unfavorable regions and
we obtain a $5\sigma$ hierarchy sensitivity for all the values of true $\delta_{CP}$. 
However for the combination of T2HK+HK+DUNE, its possible to get a hierarchy sensitivity as high as $15 \sigma$ for all the values of $\delta_{CP}$ for both the hierarchies.
In the right panel of Fig. \ref{fig1}, we have shown the hierarchy sensitivity of HK alone. From that plot we understand that hierarchy sensitivity of HK is poor if the true hierarchy is IH.
This feature can be explained from the bottom row of Fig. \ref{fig1}, where we have plotted the total electron (left panel) and the 
total muon (right panel) events as a function of $\delta_{CP}$ for different values of
$\theta_{23}$. From the plot we notice that in both the panels NH-LO is degenerate with IH-HO. Thus because of this degeneracy the hierarchy sensitivity of HK for IH is poor as compared to NH.

\subsection{Octant and CP}

\begin{figure}[ht!]
%\hspace{-0.4 in}
\includegraphics[width=0.6\linewidth]{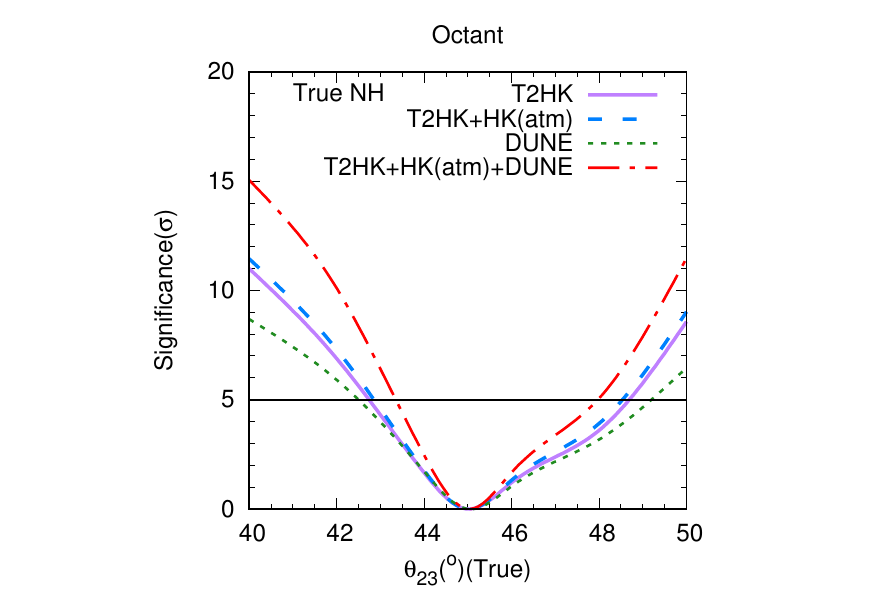}
\hspace{-1.4 in}
\includegraphics[width=0.6\linewidth]{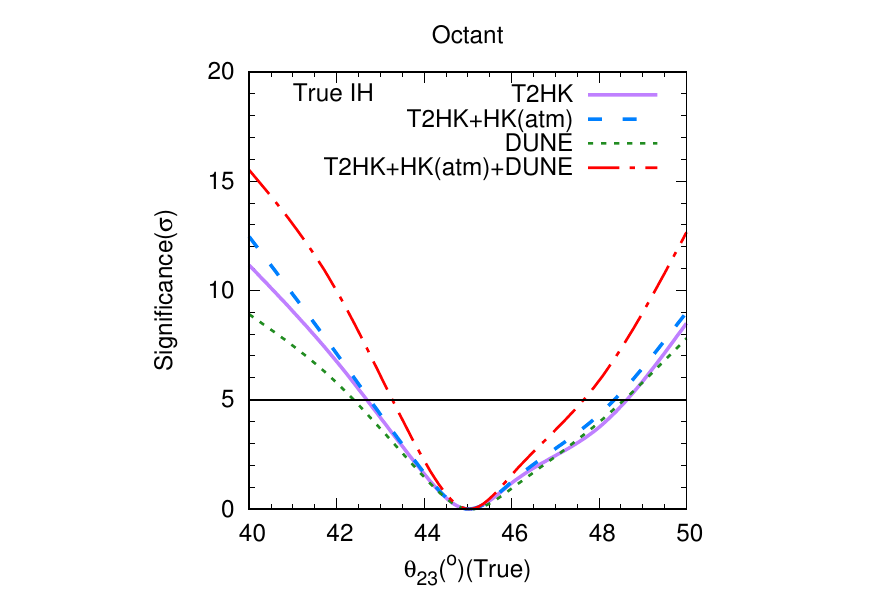} \\
\hspace{-0.7 in}
\includegraphics[width=0.6\linewidth]{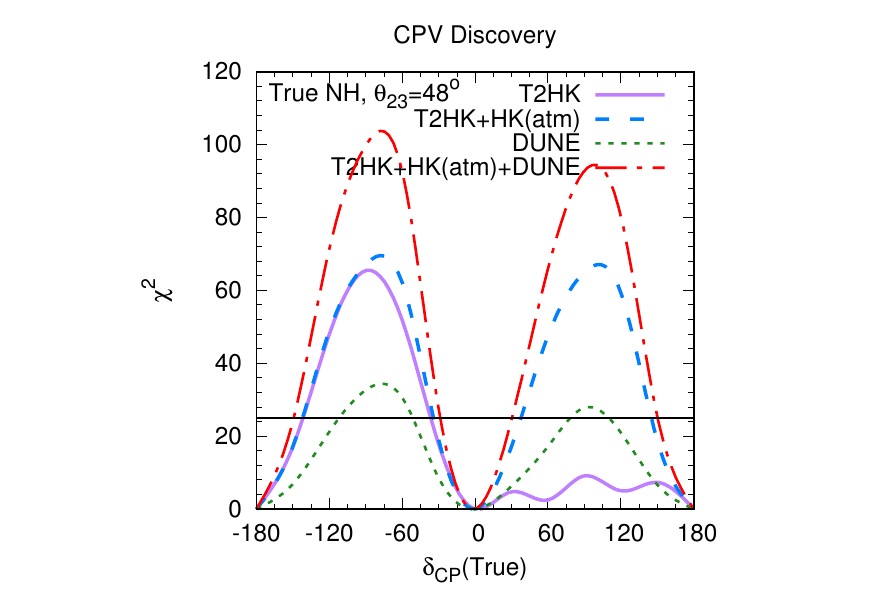}
\hspace{-1.4 in}
\includegraphics[width=0.6\linewidth]{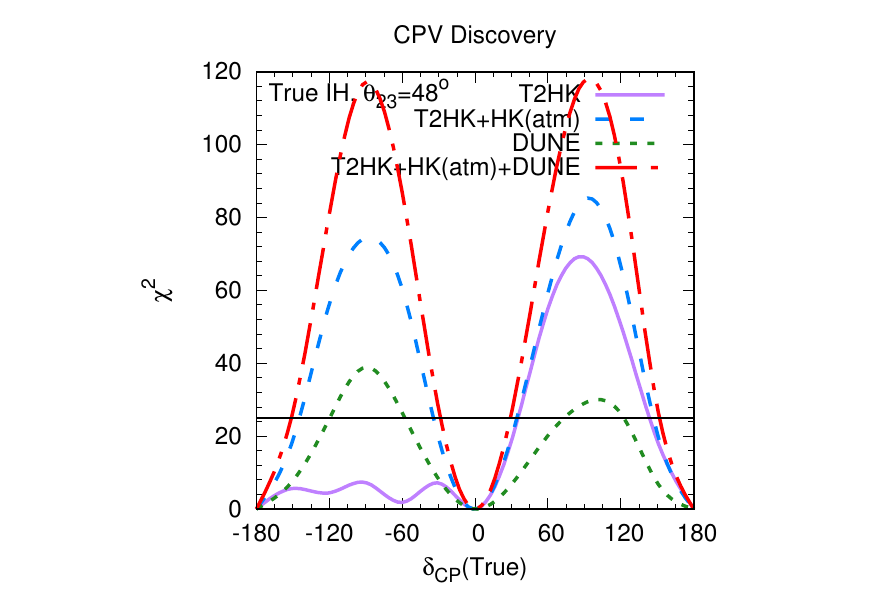} \\
\hspace{-0.4 in}
\includegraphics[width=0.6\linewidth]{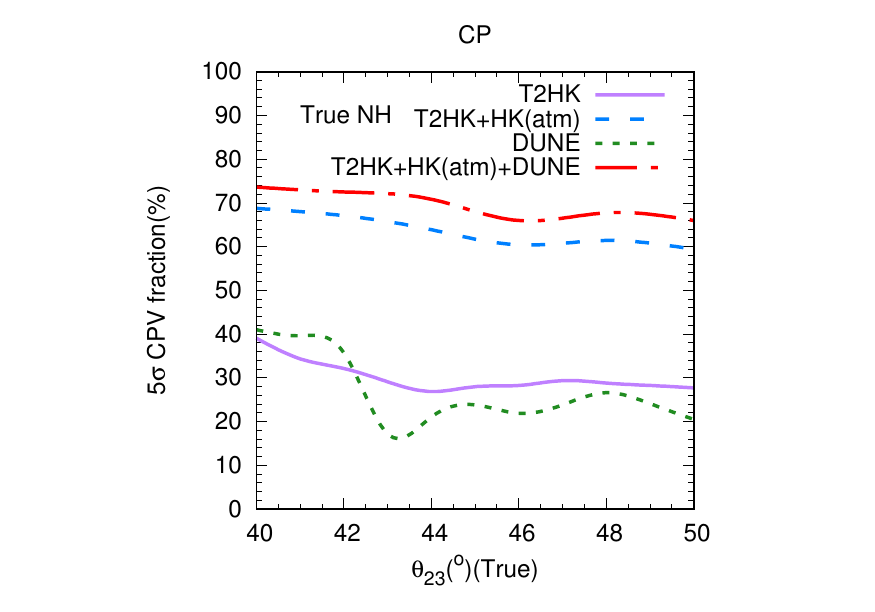}
\hspace{-1.4 in}
\includegraphics[width=0.6\linewidth]{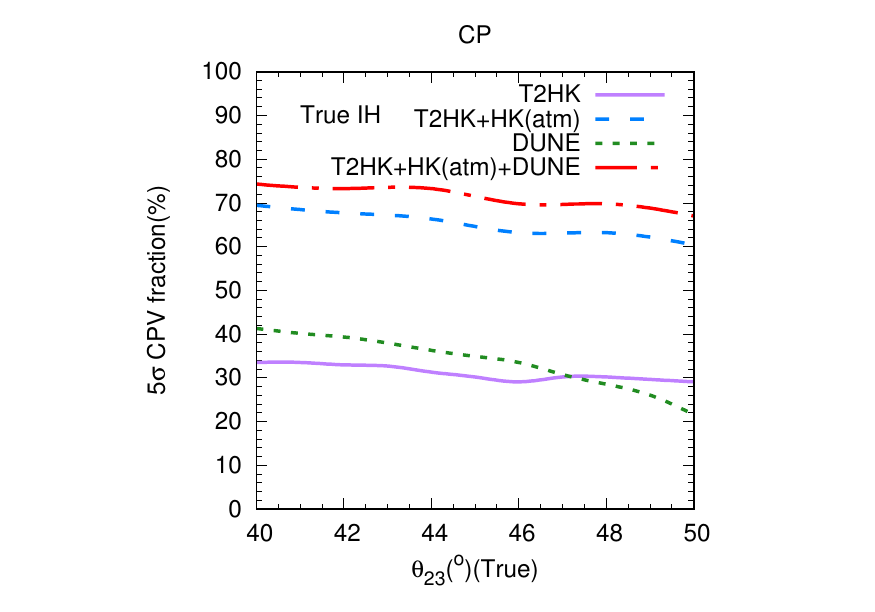}
\caption{Octant and CP sensitivity}
\label{fig2}
\end{figure}

To calculate octant sensitivity $\chi^2$ we have taken the true octant in the simulated data and the wrong octant in the theory. In this process we have marginalized over $\delta_{CP}$ and 
sign of $\Delta m_{31}^2$ in theory. We have presented our results for octant sensitivity in the upper row of Fig. \ref{fig2} as a function of true $\theta_{23}$.
From the figures we see that the octant sensitivity of T2HK, T2HK+HK and DUNE to exclude wrong octant solution at $5\sigma$ C.L is almost same. However, for the combined setup i.e., T2HK+HK+DUNE, there is a
significant amount of improvement in the octant sensitivity. For the combined setup it is possible to have a $5 \sigma$ octant sensitivity except $43.5^\circ < \theta_{23} < 48^\circ$ for both the hierarchies.

In the bottom row of Fig. \ref{fig2}, we have presented the CP violation (CPV) discovery $\chi^2$ as a function of true $\delta_{CP}$. The CPV discovery potential of an experiment is defined 
by its capability to distinguish a true value of $\delta_{CP}$ other than $0^\circ$ and $180^\circ$. In calculating CPV $\chi^2$ we have marginalized over $\theta_{23}$ and sign of $\Delta m_{31}^2$ in theory.
We have presented our results for $\theta_{23}=48^\circ$ which is the current best-fit value of this mixing angle as obtained from the global analysis of the world data.
From the plots we see that CP sensitivity of T2HK is poor in the unfavorable parameter space as mentioned earlier. But if we add HK data to T2HK, then 
there is an improvement in the unfavorable region. This is because the hierarchy sensitivity of HK resolves the hierarchy-$\delta_{CP}$ degeneracy. For the combined setup we find that a $10\sigma$ CP sensitivity
is obtained for $\delta_{CP} = \pm 90^\circ$. In the lower panel of Fig. \ref{fig2} we have plotted the fraction of $\delta_{CP}$ for which a $5\sigma$ CPV can be discovered as a function of $\theta_{23}$.
From these figures wee see that the sensitivity of T2HK and DUNE is similar for both the hierarchies. From the panels we also note that for the combined setup i.e., T2HK+HK+DUNE, CPV can be discovered at
$5\sigma$ for at least 70\% values of true $\delta_{CP}$ irrespective of the true value of $\theta_{23}$.

\section{Precision of $\delta_{CP}$, $\theta_{23}$ and $\Delta m_{31}^2$}

\begin{figure}[ht!]
\includegraphics[width=0.6\linewidth]{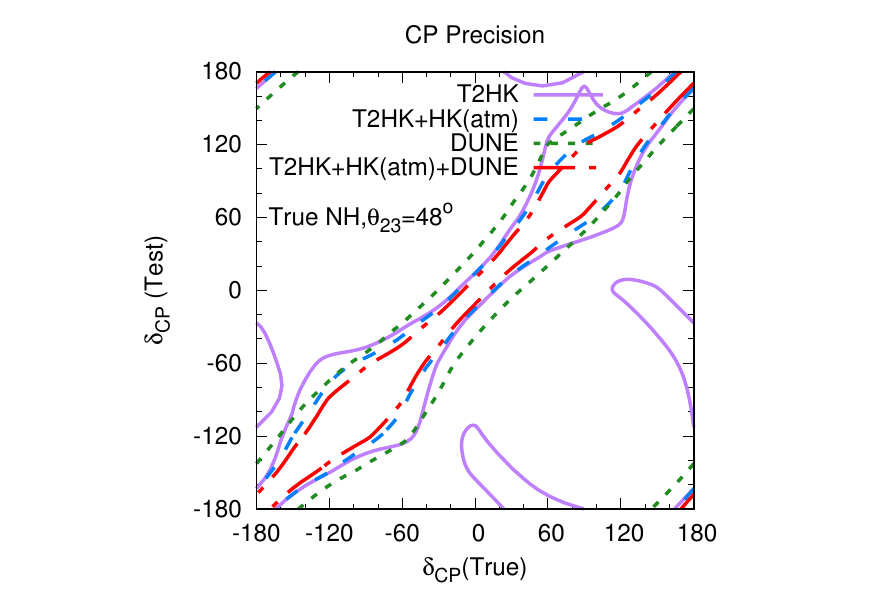}
\hspace{-1.4 in}
\includegraphics[width=0.6\linewidth]{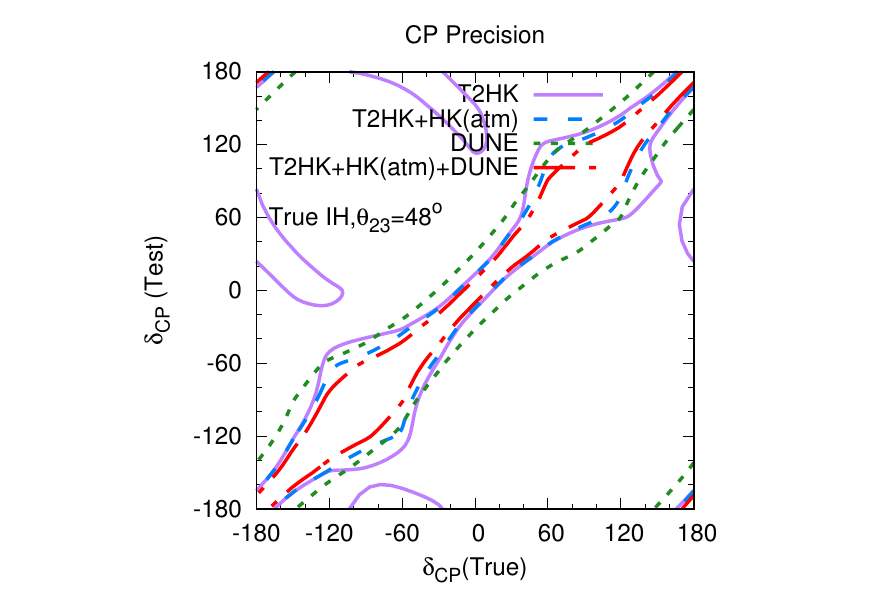} \\
\includegraphics[width=1.0\linewidth]{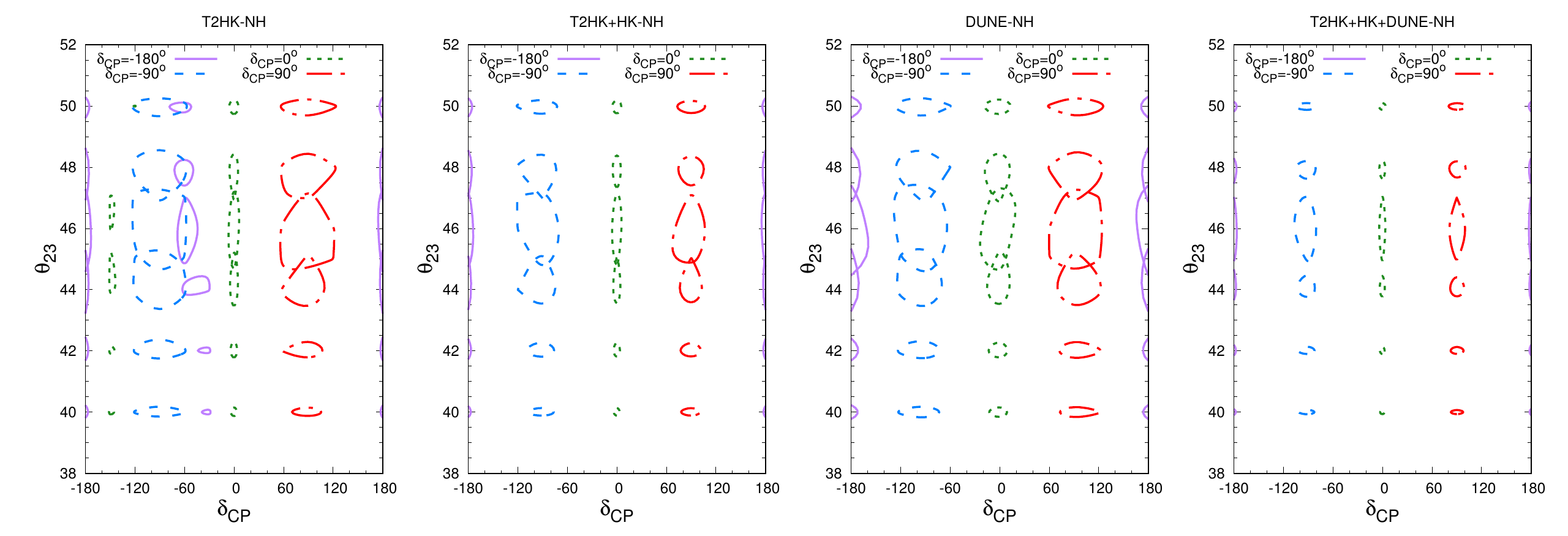} \\
\includegraphics[width=1.0\linewidth]{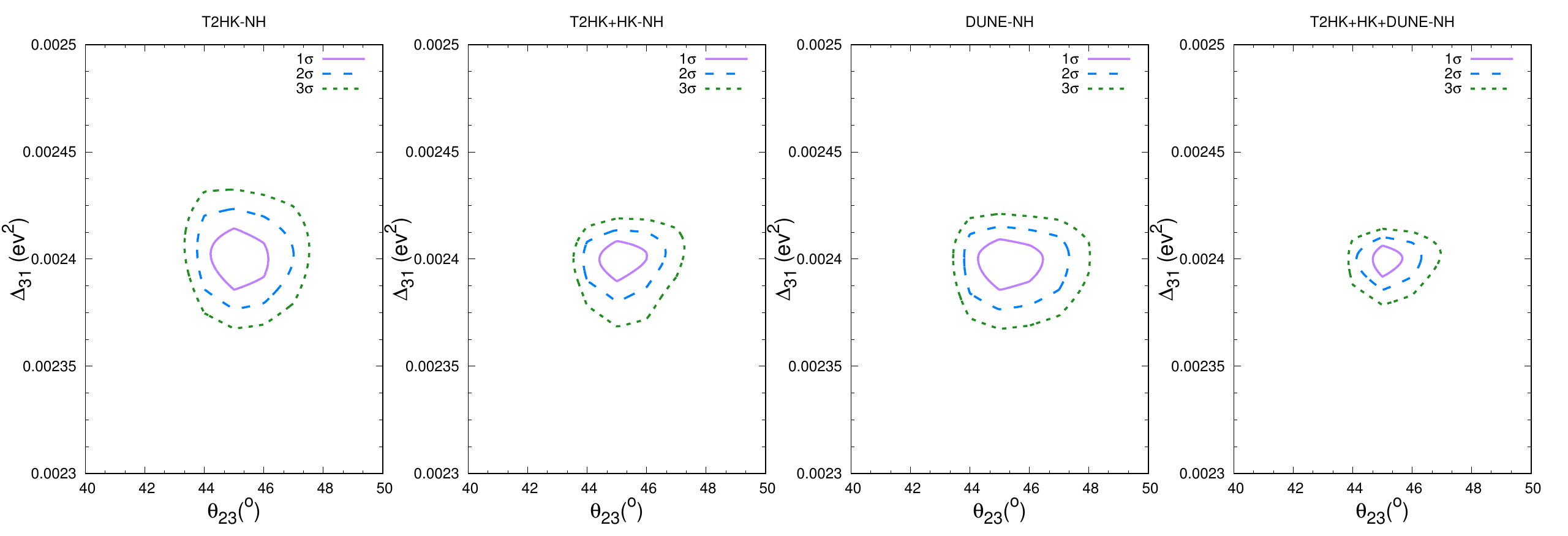}
\caption{Precision of $\delta_{CP}$, $\theta_{23}$ and $\Delta m_{31}^2$}
\label{fig3}
\end{figure}

To gauge the precision of $\delta_{CP}$ in this setup, in the upper panels of Fig. \ref{fig3} we have given $3 \sigma$ C.L contours in the true $\delta_{CP}$ - test $\delta_{CP}$ plane. We have given our results for
$\theta_{23}=48^\circ$. From the plots we see that the precision of $\delta_{CP}$ is poor for T2HK. Apart from the true diagonal solution, there are also off diagonal fake solutions. As mentioned earlier, these
wrong solutions appear due to the hierarchy-$\delta_{CP}$ degeneracy. But if we add HK data, then we see that these fake solutions disappears. This is again due to the hierarchy sensitivity of HK. 
We also notice from the figures that as we keep adding the data from different experiments, the precision of $\delta_{CP}$ gets improved.
In the middle row we have given the 90\% precision plots in the test $\delta_{CP}$ - test $\theta_{23}$ plane for different true values of $\delta_{CP}$ and $\theta_{23}$. Here we have presented our results for
normal hierarchy. From the plot we see that for true $\delta_{CP}=-180^\circ$, there are wrong CP solutions for T2HK around $\delta_{CP} = -30^\circ$. But when HK data is added to it we see that the wrong
CP solutions get vanished. For the combined setup i.e., T2HK+HK+DUNE, we find that $\delta_{CP}$ can be measured within 20\% precision for the true value of $\delta_{CP} = -90^\circ$ and $\theta_{23} = 46^\circ$.
In the bottom row of Fig. \ref{fig3}, we have plotted the 1, 2 and 3 $\sigma$ precision plots in $\theta_{23}$ - $\Delta m^2_{31}$ plane. We show our results only for NH. The results for IH is similar as that of NH.
From the plots we see that the sensitivity of the combined setup is always better the individual set up. For the combines setup of T2HK+HK+DUNE, it is possible to attain a 0.4\% precision in $\Delta m^2_{31}$ 
and a 1.3\% precision for $\theta_{23}$.

\section{Conclusion}
In this talk we have presented the sensitivity
of T2HK, HK and DUNE to mass hierarchy, octant of the mixing angle $\theta_{23}$ and $\delta_{CP}$.
Although it is difficult for T2HK to resolve
the sign degeneracy for unfavorable region
of the CP phase, when we combine it with
the atmospheric neutrino measurement at
Hyperkamiokande, we can determine the
mass hierarchy at 5\,$\sigma$ C.L. for any
value of $\delta_{CP}$. 
On the other hand, DUNE can determine the
mass hierarchy at least at 8\,$\sigma$ C.L.
by itself.  Furthermore, if we combine
all of them, then the significance to
mass hierarchy is at least 15\,$\sigma$ C.L.
In our analysis we found out that the octant sensitivity of T2HK, T2HK+HK and DUNE are quite similar in ruling out the wrong octant at $5 \sigma$ C.L.
But for T2HK+HK+DUNE the increase in the octant sensitivity is significant.
For CP violation discovery we find that the combination T2HK+HK
can measure
CP violation at 8\,$\sigma$ C.L. for
$\delta_{CP}=\pm 90^\circ$ and for T2HK+HK+DUNE
the significance for
CP violation is around 10\,$\sigma$ C.L.
for $\delta_{CP}=\pm 90^\circ$.
It is also quite impressive that with the combination of all the three experiment CP violation can be established at $5 \sigma$ C.L for at least $70\%$ true values of $\delta_{CP}$.
In the combination of all these
experiments above, the precision in $\Delta m^2_{31}$ and
$\theta_{23}$ is 0.4\% and 1.3\%,
which is an improvement by one order of magnitude
in precision with respect to the current data.
On the other hand,
the precision in $\delta_{CP}$ is 20\%.
We will be in the era of
precision measurements of neutrino
oscillation parameters, and
combination of Hyperkamiokande and DUNE
will play an important role in determination
of $\delta_{CP}$ as well as $\theta_{23}$.

\section*{Acknowledgements}
This research was partly supported by a Grant-in-Aid for Scientific
Research of the Ministry of Education, Science and Culture, under
Grants No. 25105009, No. 15K05058, No. 25105001 and No. 15K21734.

\end{document}